\documentclass{aa}  

\usepackage{graphicx}
\usepackage{txfonts}
\usepackage{hyperref}
\usepackage{natbib}
\hypersetup{
    colorlinks=true,
    citecolor={cyan}
}
\defcitealias{duBuisson+2020}{dB20}
\begin{document}

   \title{An upper limit on the spins of merging binary black holes formed through binary evolution}

   \author{Pablo Marchant\inst{1}
          \and
          Philipp Podsiadlowski\inst{2}
          \and
          Ilya Mandel\inst{3,4}
          }

   \institute{Institute of Astronomy, KU Leuven, Celestijnenlaan 200D, B-3001 Leuven, Belgium\\
              \email{pablo.marchant@kuleuven.be}
         \and
             University of Oxford, St Edmund Hall, Oxford OX1 4AR, UK
             \and
             School of Physics and Astronomy, Monash University, Clayton, VIC 3800, Australia
             \and
             The ARC Centre of Excellence for Gravitational Wave Discovery -- OzGrav, Australia
             }


 
  \abstract
   {As ground-based gravitational wave detectors progressively increase their sensitivity, observations of black hole mergers will provide us with the joint distribution of their masses and spins. This will be a critical benchmark to validate different formation scenarios.}
   {Merging binary black holes formed through the evolution of isolated binary systems require both components to be stripped of their hydrogen envelopes before core-collapse. The rotation rates of such stripped stars are constrained by the critical rotation limit at their surface, setting a restriction on their angular momentum content at core-collapse. We aim to determine if this restriction plays a role in the spins of binary black hole mergers.}
   {We use detailed calculations of stripped stars with the \texttt{MESA} code at low metallicities ($Z=Z_\odot/10$, $Z_\odot/50$ and $Z_\odot/250$) to determine the dimensionless spins of black holes produced by critically rotating stellar progenitors. To study how such progenitors can arise, we consider their formation through chemically homogeneous evolution (CHE) in binary stars. We use a semi-analytical model to study the physical processes that determine the final angular momentum of CHE binaries, and compare our results against available population synthesis models that rely on detailed binary evolution calculations.}
   {We find that above black hole masses of $\simeq 25M_\odot$, the dimensionless spin parameter of critically rotating stripped stars ($a=Jc/(GM^2)$) is below unity. This results in an exclusion region at high chirp masses and effective spins that cannot be populated by isolated binary evolution. CHE can produce binaries where both black holes hit this limit, producing a pile-up at the boundary of the excluded region. Highly spinning black holes arise from very low-metallicity CHE systems with short delay times, which merge at higher redshifts. On the other hand, the contribution of CHE to merging binary black holes detected in the third observing run of the LVK collaboration is expected to be dominated by systems with low spins ($\chi_\mathrm{eff}<0.5$) which merge near redshift zero. Owing to its higher projected sensitivity and runtime, the fourth observing run of the LVK collaboration can potentially place constraints on the high spin population and the existence of a limit set by critical rotation.}
   {}


   \maketitle
%

\section{Introduction}

Ever since the first observation of a merging binary black hole (BBH) \citep{Abbott_GW150914_2016}, continuous improvements to ground-based interferometers have resulted in weekly discoveries when current generation detectors are online. As detector biases are well understood, with close to 100 detected sources \citep{GWTC1,GWTC2,GWTC3, Nitz+2023} it has become possible to place constraints on the intrinsic distribution of merging BBH masses and spins at redshifts near zero \citep{LIGOpop3}. Since the number of sources is still limited the inference of intrinsic distributions mostly relies on parameterized models, including features such as a bump or cutoff in the mass distribution which is expected from pulsational pair-instability supernovae (PPISNe, \citealt{Belczynski+2014, Marchant+2016, Belczynski+2016b, Woosley2017, FishbachHolz2017, TalbotThrane2018, Marchant+2019, Farmer+2019}). Different parameterized distributions can, however, lead to physically different interpretations. The model used by the LIGO/Virgo/KAGRA collaboration for the distribution of black hole spins indicated counter-aligned black hole spins \citep{LIGOpop3}, while \citet{Roulet+2021,Galaudage:2021,Callister+2021,Hoy:2021} argued that this conclusion is dependent on the model distribution used.

With an increasing number of measurements it is also becoming possible to determine correlations between different BBH properties and to place constraints on their joint distributions. For low-mass binaries the best constrained intrinsic property is the chirp mass,
\begin{eqnarray}
M_\mathrm{chirp}=\frac{(m_1 m_2)^{3/5}}{(m_1+m_2)^{1/5}},
\end{eqnarray}
where $m_1$ and $m_2$ are the individual component masses. Useful constraints are also placed on both the effective spin $\chi_\mathrm{eff}$ and mass ratio $q$, defined as
\begin{eqnarray}
\chi_\text{eff}\equiv \frac{m_1 a_1\cos\theta_1 + m_2 a_2\cos\theta_2}{m_1+m_2},\quad q=\frac{m_2}{m_1}
\end{eqnarray}
where $a_1$ and $a_2$ are the dimensionless spin parameters of each BH while $\theta_1$ and $\theta_2$ are the angles between the individual spin vectors and the orbital angular momentum. Despite there being a significant correlated uncertainty between $\chi_\mathrm{eff}$ and $q$ for individual detections \citep[e.g.,][]{Hannam+2013,MandelSmith:2021}, current observations already suggest the presence of an intrinsic (anti)correlation between these two properties, as well as a non trivial dependence of the $\chi_\mathrm{eff}$ distribution with respect to $M_\mathrm{chirp}$ \citep{Safarzadeh+2020,Callister+2021,LIGOpop3, FrancioliniPani2022,Adamcewicz:2023}. Our understanding of these distributions will rapidly evolve, as the currently ongoing O4 observing run is expected to significantly increase the number of observed BBH mergers \citep{Abbott+2020_prospects}.

The precise distribution of spins is critical to differentiate between various formation scenarios of merging BBHs. Dynamical processes in dense environments can result in an isotropic distribution of angles between the orbital plane and the spins of each BH in contrast to near-aligned spins in binary formation channels \citep{MandelOShaughnessy2010, Rodriguez+2016b}. High black hole spins can also be indicative of hierarchichal BH mergers, as a generic outcome from the merger of two similar-mass slowly-spinning BHs is the formation of a BH with a dimensionless spin $\sim0.7$ \citep{Pretorius2005, BertiVolonteri2008,GerosaBerti2017,Rodriguez+2019, Hiromichi+2021, GerosaFishbach2021}. In isolated binary systems it has been predicted that envelope stripping coupled with efficient angular momentum transport between the stellar core and the outer layers should lead to small aligned BH spins \citep{Belczynski+2020}.

Merging BBHs with high spins ($\chi_\mathrm{eff}>0.5$) could also arise in binary systems, either through tidal coupling in a short period post-common-envelope binary \citep{Qin+2018, Zaldarriaga+2018, Bavera+2020,Bavera+2021, FullerLu2022, MaFuller2023} or through chemically homogeneous evolution (CHE).  Large- scale flows are predicted to mix the hydrogen burning core with the envelope in rapidly rotating massive stars, preventing radial expansion in the CHE channel  \citep{Maeder1987}.  \citet{deMink+2009} argued that this process operates in short period binaries (periods of the order of one day) where rapid rotation arises from tidal synchronization. If both components of a massive binary undergo CHE, it is then possible to form a merging BBH without the need of an additional process to harden the binary \citep{MandeldeMink2016,Marchant+2016,deMinkMandel2016,duBuisson+2020,Riley+2021,StevensonClarke2022}. BBHs formed through CHE can then produce pairs of high spin BHs \citep{Marchant+2016, FullerMa2019}, although this is sensitive to mass loss after the main sequence.

Various attempts have been made to constraint the individual contributions of different formation channels to the observed sample of merging compact objects. One possible approach to this is the use of individual predicted distributions from different channels, which can then be scaled to attempt to reproduce the observed distribution coupled with its observation biases \citep{Zevin+2017,Bouffanais+2019,Safarzadeh+2020,Wong+2021, Zevin+2021,Perigois+2023}. Alternatively, predicted distributions from each channel can be used to infer individually what is the likelihood that a single source is produced by a specific formation scenario (e.g. \citealt{Qin+2022, Antonelli+2023}). However, conclusions from model comparison can be severely affected if one lacks knowledge of some formation channels \citep{QiuCheng+2023}, and predicted distributions have large uncertainties with different codes predicting drastically different outcomes even for individual binary systems (e.g. \citealt{Belczynski+2022}). It is then critical to identify the qualitative features that arise from specific formation channels both in masses and spins, independent of large uncertainties in population modelling (e.g. \citealt{vanSon+2022}).

In this work we study the limit set by the critical rotation rate of a BH progenitor on its resulting spin. In Section \ref{test} we discuss the physics of this limit and how it is expected to scale with stellar mass. In Section \ref{sec:semi} we connect this limit with evolutionary processes to form merging BBHs, by developing a simple semi-analytical model that describes the evolution of CHE binaries. We then compare our results to the detailed population synthesis calculations of \citet{duBuisson+2020} (hereafter \citetalias{duBuisson+2020}), which we have re-analyzed to include predictions on the spins of BBHs produced via CHE. We conclude our work and discuss its broader implications in Section \ref{sec:conclusion}.

\section{Stellar critical rotation and black hole spins} \label{test}
In the past two decades, various results have pointed to the need of efficient angular momentum transport in stellar interiors. This includes asteroseismic measurements of the rotation rates of the cores of evolved stars (see \citealt{Aerts+2019} for a recent review) as well as the angular momentum content of white dwarfs and young neutron stars \citep{Suijs+2008}. One potential mechanism that allows for efficient angular momentum transport throughout the life of a star is the Tayler-Spruit (TS) dynamo \citep{Spruit2002}. This dynamo process amplifies seed magnetic fields through differential rotation in radiative layers, leading to a stronger coupling of the stellar core and its outer layers. \citet{Fuller+2019} further developed the model for the TS dynamo, and argued this process is more efficient than previously thought (we refer to this model as the TSF dynamo). The model by \citet{Fuller+2019} provides a better fit to asteroseismic measurements of red giant and red clump stars, and to the rotation rates of white dwarfs. It also leads to near solid-body rotation of the hydrogen depleted cores.

The TSF dynamo reduces significantly the final rotation rates of massive stripped stars, as it couples the outer mass-losing layers to the core that remains until core-collapse \citep{FullerMa2019}. If angular momentum transport is efficient enough to make a stripped star remain near solid-body rotation until late burning phases, the critical rotational frequency $\Omega_\mathrm{crit}$ sets a limit on the maximum angular momentum of the star,
\begin{eqnarray}
    J_\mathrm{max} = I \Omega_\mathrm{crit},
\end{eqnarray}
where $I$ is the moment of inertia. The critical rotation rate $\Omega_\mathrm{crit}$ represents the limit at which centrifugal, radiation and gravitational forces are perfectly balanced at the stellar equator. If the star collapses directly into a BH, then its dimensionless BH spin is limited to
\begin{eqnarray}
    a_\mathrm{BH} = \min\left(1,\frac{J_\mathrm{max}c}{M^2G}\right),
\end{eqnarray}
where $M$ is the stellar mass. In practice, as the BH is assembled, if some layers have enough angular momentum to support a disk at the innermost stable circular orbit then outflows can remove the surplus angular momentum and form a sub-critical BH (e.g. \citealt{BattaRamirezRuiz+2019,MurgiaBerthier+2020}).

Owing to the large luminosities of massive stars, one needs to account for the impact of radiation on the critical rotation rate. Ignoring gravity darkening, the critical rate depends on the Eddington factor $\Gamma$ and the equatorial radius $R_\mathrm{eq}$ as \citep{Langer1997}
    \begin{eqnarray}
        \Omega_\text{crit}=\sqrt{\frac{GM(\Gamma-1)}{R_\mathrm{eq}^3}},\quad \Gamma=\frac{\kappa L}{4\pi c G M},
    \end{eqnarray}
    with $\kappa$ and $L$ being the surface opacity and luminosity, respectively. The critical rotation rate including radiation is commonly referred to as the $\Gamma-\Omega$ limit. In this section we aim to provide a qualitative estimate of the impact of the $\Gamma-\Omega$ limit on BH spins.  We will use non-rotating stellar models to estimate $a_\mathrm{BH}$. Even if we were to use rotationally deformed stellar models, the expression for the critical rotation rate of \cite{Langer1997} is only approximate as it ignores the impact of gravity darkening \citep{MaederMeynet2000b,Sanyal+2015,Fabry+2022}.

The dependence of the limiting $a_\mathrm{BH}$ on mass can be understood by expressing the moment of inertia as
\begin{eqnarray}
    I = kMR^2,
\end{eqnarray}
where $k$ is the square of the ratio between the gyration radius and the stellar radius. Here we ignore rotational deformation and use $R_\mathrm{eq}=R$. If we take $k$ to be independent of mass at a given evolutionary stage (which is the case for polytropic models with the same polytropic index), and consider mass-radius and mass-luminosity relationships of the form $R\propto M^{\beta}$ and $L\propto M^{\alpha}$ respectively, one has that:
\begin{eqnarray}
\frac{J_\mathrm{max}c}{M^2G} \propto M^{(\beta-1)/2}\sqrt{\Gamma-1},\quad \Gamma \propto M^{\alpha-1}.\label{eq:spinlimit}
\end{eqnarray}
Power law relationships are not valid across all stellar masses, so the power law exponents are normally also taken to be functions of mass. In particular, the mass-luminosity exponent for main sequence stars is $\alpha>1$, while it approaches unity at very high masses (e.g. \citealt{Kohler+2015}). This leads to $\Gamma\rightarrow 1$ as mass increases, which pushes down the spin limit in Equation (\ref{eq:spinlimit}). Similarly, both stellar models and homology relationships show that main sequence massive stars have $\beta<1$ \citep{Kippenhahn+2013}, which is also found from observations of eclipsing binaries (e.g. \citealt{Torres+2010}, \citealt{Eker+2018}). Having $\beta<1$ indicates that the spin limit in Equation (\ref{eq:spinlimit}) is reduced with mass. As more massive stars approach the Eddington limit near their surface they might experience inflation, where a small amount of mass expands to large radii (e.g. \citealt{Sanyal+2015}). In this case $\beta>1$ but also the approximation that $k$ is independent of mass is not valid . If we want to understand in detail how the critical rotation limit on spins works at various evolutionary stages and masses, we need to rely on numerical calculations.

To make quantitative estimates we compute mock models of stripped stars using version \texttt{23.05.1} of the \texttt{MESA} stellar evolution code\footnote{Input files to reproduce our simulations together with our data are available at \url{https://zenodo.org/doi/10.5281/zenodo.10199794}.} \citep{Paxton+2011,Paxton+2013,Paxton+2015,Paxton+2018,Paxton+2019,Jermyn+2023}. Details about input microphysics are described in Appendix \ref{appendix::mesa}. We will also use these models to discuss the spins produced by CHE stars, so evolution is initiated at the zero-age main-sequence (ZAMS). We consider non-rotating models without mass loss which are forced to evolve homogeneously until core helium depletion by enforcing a large diffusion mixing coefficient throughout the entire star.
Initial masses are sampled logarithmically between $\log_{10}M/M_\odot=0.8$ ($6.3M_\odot$) and $\log_{10}M/M_\odot=2.4$ ($251M_\odot$) and kept fixed during the evolution. Three metallicites are sampled, $Z_\odot/10$, $Z_\odot/50$ and $Z_\odot/250$ where the solar metallicity $Z_\odot=0.0142$ and the relative metal fractions are adopted from \cite{Asplund+2009}. All simulations are run until core carbon depletion.

As we do not include mass loss, the resulting models are not meant to represent the evolution of a single star, but to be indicative of the structure of a star with a given mass at a specific evolutionary stage. Since CHE stars in binaries contract and tidally decouple after the main sequence, stellar winds can efficiently spin them down and we do not expect them to remain homogeneous. However, a stellar model evolved without mass loss will produce a composition discontinuity between the CO core and the remaining helium envelope, which is not present in mass-losing models. It is for this reason that we keep our models homogeneous through helium burning, to better represent the structure of a model at each evolutionary stage and metallicity.

\begin{figure}
\includegraphics[width=\columnwidth]{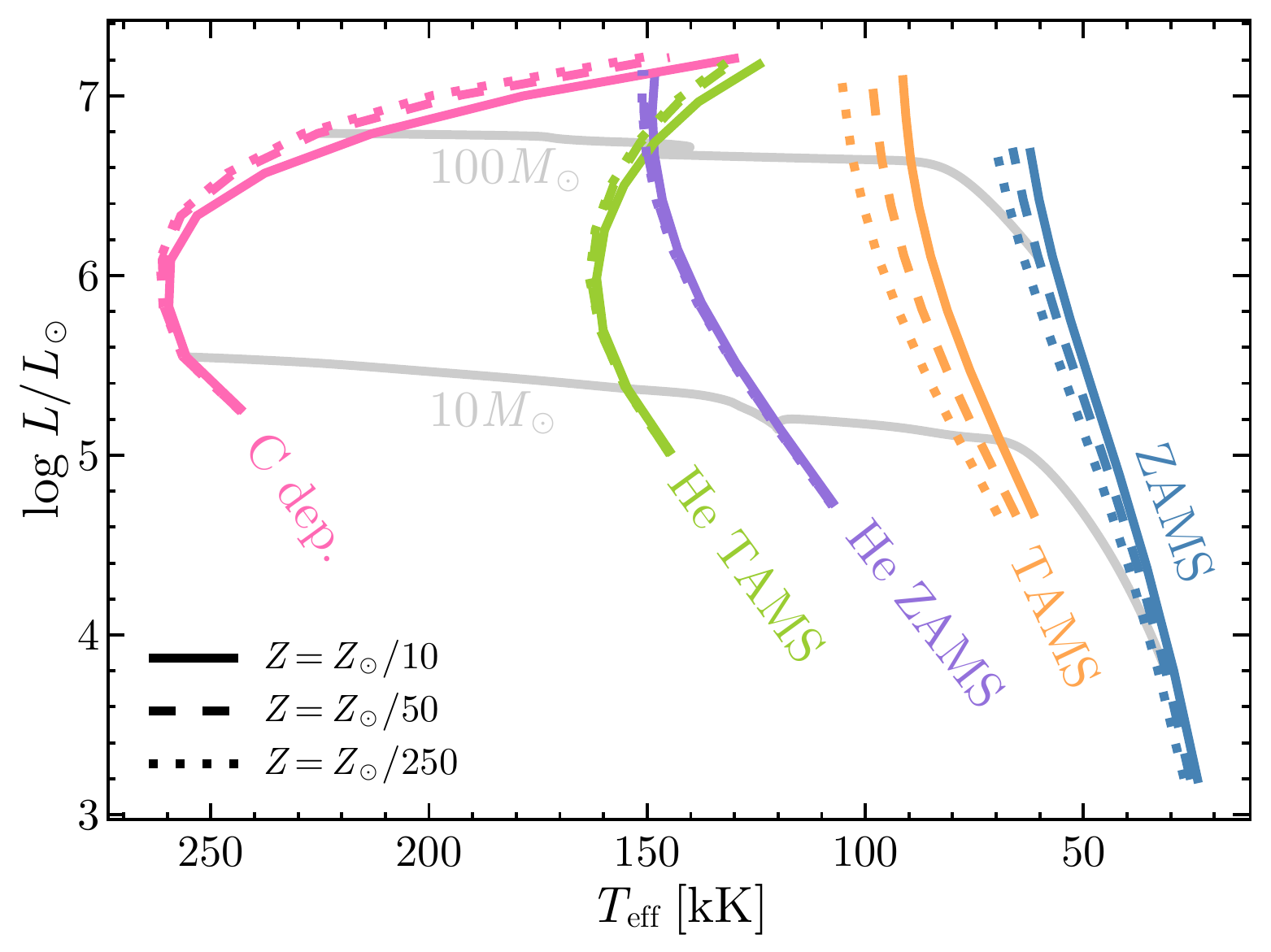}
\caption{The mock CHE models described in section \ref{test} on the Hertzsprung-Russell diagram. Colors indicate different evolutionary stages including the zero-age main sequence (ZAMS), the terminal age main sequence (TAMS), the helium ZAMS and TAMS, and core carbon depletion (C dep.). Different linestyles indicate different values for the metallicity. Two evolutionary tracks with a metallicity of $Z_\odot/50$ and masses of $10M_\odot$ and $100M_\odot$ are shown in gray for reference.}\label{fig:HR}
\end{figure}

\begin{figure}
\includegraphics[width=\columnwidth]{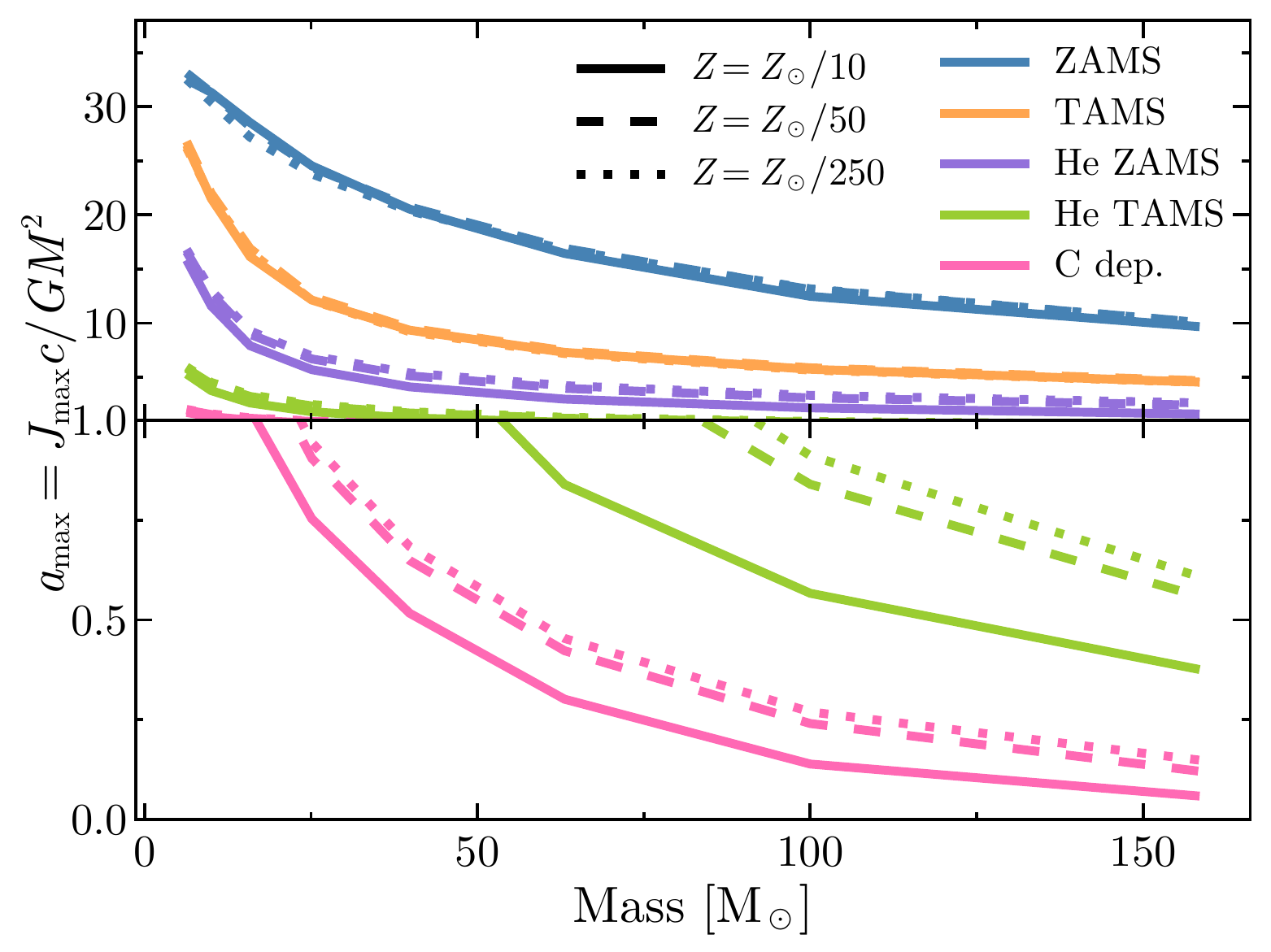}
\caption{Dimensionless spins corresponding to the angular momentum and mass of a critically rotating star undergoing CHE at different evolutionary stages. Values are estimated from the non-rotating models described in Section \ref{test}.}\label{fig:maxspin}
\end{figure}

The position in the Hertzsprung-Russell diagram for the mock CHE models at different evolutionary stages is illustrated in Figure \ref{fig:HR}. As the stars evolve, they continuously increase their luminosity and (except for some of the most massive models) continuously contract. At each of these stages we can consider the maximum angular momentum that the star can contain under solid-body rotation, $J_\mathrm{max}$, and compute the resulting maximum spin $a_\text{max}=J_\text{max}c/GM^2$ a black hole would have were the star to collapse at that stage. As shown in Figure \ref{fig:maxspin} we find that for all metallicities and masses $a_\text{max}$ decreases as the star evolves. In particular, at core-carbon depletion, stars with masses above $25M_\odot$ would not be able to produce a maximally spinning BH. Barring the assumption of solid-body rotation, which we discuss later, this is a feature that should be common not only to CHE but to any formation scenario that involves the formation of BHs through stripped envelope stars. This also includes stripped stars that are tidally spun up by their companion (e.g. \citealt{Qin+2018}). A fit to $a_\mathrm{max}$ at core carbon depletion is provided in Appendix \ref{app:spin}.

The limit we find on BH spins is consistent with currently measured values in X-ray binaries. The most massive BH with a dynamical mass estimate and a constrained spin is Cyg X-1, with a mass of $21.2\pm 2.2M_\odot$ and $a>0.9985$ \citep{MillerJones+2021}. It might be possible in the near future to identify more massive BHs in X-ray binaries, which would allow to test our predicted spin limit with electromagnetic observations. On the other hand, merging binary BHs with masses $>20M_\odot$ are now frequently detected, possibly allowing a direct test. Even if critical rotation sets a limit on the spins of BHs formed from stripped stars, for this spin limit to appear in GW observations we require that the evolutionary processes which form merging binary BHs also lead to critically rotating stars close to core-collapse. In the context of long gamma-ray burst (lGRB) formation via the collapsar model \citep{Woosley1993}, tidal spin-up of stripped stars by degenerate companions has been suggested as a mechanism to form a rapidly spinning BH \citep{Podsiadlowski+2004,Izzard+2004}. If such systems also result in a compact object merger, then the second formed compact object would dominate the effective spin of the system \citep{Qin+2018, Bavera+2020, Bavera+2021}. Another mechanism to form high spin merging binary BHs is the CHE scenario \citep{MandeldeMink2016, deMinkMandel2016, Marchant+2016}. In the next section we consider the CHE scenario, by means of a simple semi-analytical model that lets us understand what limits the masses and spins that can be produced by this channel.

\section{Semi-analytical model for chemically homogeneous evolution}\label{sec:semi}
To explore the possible parameter space in chirp mass and effective spin covered by CHE, we construct a simple semi-analytical model. This model relies on a few physical quantities obtained from stellar evolution models described in Section \ref{test} and provides clear insight into the physics that determine the boundaries in BBH properties produced by CHE. For simplicity, we consider only equal-mass binaries and assume solid-body rotation. Through multiple evolutionary stages we need to account for the evolution of the angular momentum of each component, as well as their masses and orbital periods. We describe our assumptions at each stage next.

\begin{enumerate}
\item TAMS: We ignore evolution during core hydrogen burning and assume as a starting point that both stars have evolved chemically homogeneous and have reached the TAMS while remaining tidally synchronized. The parameters that describe this stage are the orbital period $P_\mathrm{TAMS}$ and the mass of each star $M_\mathrm{TAMS}$. Given the assumption of synchronicity the angular momentum of each star at this stage is equal to
\begin{eqnarray}
    J_\mathrm{TAMS} = \frac{2\pi}{P_\mathrm{TAMS}}I_\mathrm{TAMS}.
\end{eqnarray}
The orbital period at TAMS cannot be arbitrarily small. We consider as a limit the period at which the stars would be in contact\footnote{As a less restrictive constraint one could consider the limit at which a contact binary at TAMS would be just below the limit of L2 overflow.}. For a star of mass $M_1$ with a companion of mass $M_2$ this period can be computed from Kepler's third law and the separation at contact,
\begin{eqnarray}
    P_\mathrm{contact}=\sqrt{\frac{4\pi^2}{G(M_1+M_2)}a_\mathrm{contact}^3}.
\end{eqnarray}
The separation at contact is determined using the fit for the Roche lobe size of \cite{Eggleton1983} and the stellar radius of the CHE star at TAMS,
\begin{eqnarray}
    a_\mathrm{contact}=\frac{0.6q^{2/3}+\ln(1+q^{1/3})}{0.49q^{2/3}}R_\mathrm{1,TAMS},\quad q\equiv \frac{M_1}{M_2}.
\end{eqnarray}
\item He-ZAMS: After the main sequence, a CHE star will contract until it reaches thermal equilibrium at the beginning of core-helium burning, the He-ZAMS. We assume tidal synchronization stops during this short contraction phase and that the angular frequency of the star evolves by conserving the angular momentum at TAMS. If contraction is homologous, meaning the radial position of each mass element contracts by the same fraction $\alpha$, then the angular frequency of the star increases as $\Omega\propto\alpha^{-2}$. In contrast, ignoring changes in surface opacity or luminosity, the critical rotation rate increases as $\Omega_\mathrm{crit}\propto\alpha^{-3/2}$. The contracting star is then expected to evolve closer to critical rotation ($\Omega/\Omega_\mathrm{crit}\propto \alpha^{-1/2}$). In practice contraction is not exactly homologous, and the critical rotation rate will be modified as the luminosity and surface opacity change, but this illustrates that during contraction after TAMS stars can evolve towards critical rotation.

To determine the angular momentum at the He-ZAMS we then need to account for the maximum angular momentum $J_\mathrm{He-ZAMS,max}$ that the star can possibly have while rotating at its critical speed. We simply limit the angular momentum at this stage to this value,
\begin{eqnarray}
    J_\mathrm{He-ZAMS} = \min(J_\mathrm{TAMS}, J_\text{He-ZAMS, max}).
\end{eqnarray}
The loss of excess angular momentum needs to be accompanied by mass loss. However, if ejection happens through the formation of a decretion disk (similar to Be-stars, e.g. \citealt{Rivinius+2013}) then significant amounts of angular momentum can be removed with a small amount of mass ejected. We ignore the uncertain mass loss and orbital period evolution during the contraction from TAMS to He-ZAMS,
\begin{eqnarray}
    M_\mathrm{He-ZAMS}=M_\mathrm{TAMS},\quad P_\mathrm{He-ZAMS}=P_\mathrm{TAMS}.
\end{eqnarray}
\item Core helium burning: During the post-MS evolution the components will undergo mass loss, possibly as Wolf-Rayet stars. We parameterize the total fraction of mass lost by a free parameter $f_\mathrm{WR}$ defined as
\begin{eqnarray}
f_\mathrm{WR}\equiv \frac{M_\mathrm{He-TAMS}}{M_\mathrm{He-ZAMS}}.
\end{eqnarray}
This is such that $f_\mathrm{WR}=1$ represents no mass loss, while $f_\mathrm{WR}=0$ would indicate that the entire stellar mass is removed.

As mass is lost the orbital period of the binary will also change. Under the approximation where the stellar winds remove orbital angular momentum equal to the specific orbital angular momentum of the corresponding component, the orbital period satisfies
\begin{eqnarray}
    \frac{P_\mathrm{He-TAMS}}{P_\mathrm{He-ZAMS}}=\left(\frac{M_\mathrm{He-ZAMS}}{M_\mathrm{He-TAMS}}\right)^2= f_\mathrm{WR}^{-2}.
\end{eqnarray}
Since $0<f_\mathrm{WR}<1$, mass loss will lead to orbital expansion.

To determine the evolution of each star's angular momentum in connection to mass loss, we assume mass loss removes spin angular momentum equal to that of spherical shells of the stellar radius rotating at the surface angular frequency,
\begin{eqnarray}
\dot{J}=i_\mathrm{surf}\Omega \dot{M}, \quad i_\mathrm{surf}=\frac{2}{3}R^2.
\end{eqnarray}
Writing the moment of inertia of the star in terms of $k$ (which is the squared ratio between the gyration radius and the stellar radius) we obtain that
\begin{eqnarray}
    \frac{\dot{J}}{J} = \frac{2}{3k}\left(\frac{\dot{M}}{M}\right).
\end{eqnarray}
Our simulations indicate that during core helium burning $k$ covers the range $\sim 0.08-0.1$. Ignoring the time dependence of $k$, the previous equation can be integrated to provide the angular momentum of the star at the He-TAMS,
\begin{eqnarray}
\frac{J_\mathrm{He-TAMS}}{J_\mathrm{He-ZAMS}}=\left(\frac{M_\mathrm{He-TAMS}}{M_\mathrm{He-ZAMS}}\right)^{2/(3k)}= f_\mathrm{WR}^{2/(3k)}.
\end{eqnarray}
This can be related to the dimensionless spin of the star, i.e., the spin of a BH which a star would produce through direct collapse, $a=Jc/M^2G$. In terms of $f_\mathrm{WR}$ we have that
\begin{eqnarray}
    \frac{a_\mathrm{He-TAMS}}{a_\mathrm{He-ZAMS}} = f_\mathrm{WR}^{-2+2/(3k)}.
\end{eqnarray}
This equation indicates that the final compact object spin is reduced under our assumption of solid-body rotation as long as $k<1/3$. Owing to the central concentration of mass in stars, this is generally the case.
\item Carbon depletion and BBH formation: Similar to the evolution from TAMS to He-ZAMS, we limit the angular momentum of a star to its maximum value at critical rotation,
\begin{eqnarray}
    J_\text{C. dep}=\min(J_\mathrm{He-TAMS}, J_\text{C. dep, max}),
\end{eqnarray}
and we keep the mass and orbital period from the He-TAMS,
\begin{eqnarray}
    M_\text{C. dep}=M_\mathrm{He-TAMS},\quad P_\text{C. dep}=P_\mathrm{He-TAMS}.
\end{eqnarray}

We assume direct collapse to a BH, ignoring any loss of mass or angular momentum at stages beyond carbon depletion. For the mass range we are concerned with this is supported both by theory (e.g. \citealt{Fryer1999}) as well as observations of BH binaries with massive companions such as Cyg X-1 \citep{MirabelIrapuan2003} and VFTS-243 \citep{Shenar+2022}. The resulting mass and spin of the BHs is finally computed as
\begin{eqnarray}
    M_\mathrm{BH}=M_\text{C. dep},\quad a_\mathrm{BH}=\min\left(1,\frac{J_\text{C. dep}c}{M_\text{C. dep}^2G}\right),
\end{eqnarray}
while the orbital period of the BBH is taken to be that at carbon depletion. The delay time until BBH merger is then computed from the results of \cite{Peters1964} for the case of a circular orbit. As a simple model for the impact of (P)PISNe we ignore any BH that would form within the predicted PISN gap, taking the lower limit of the gap to be $M_\mathrm{BH}=43.94M_\odot$ from \cite{Marchant+2019}.

\end{enumerate}

Our model then relies on three input parameters: the component masses at TAMS, the orbital period at TAMS, and the fraction of mass lost after the main sequence ($M_{\mathrm{TAMS}}$, $P_\mathrm{TAMS}$ and (1-$f_\mathrm{WR}$) respectively). The model is semi-analytical as it does require results from detailed calculations of stellar evolution. Specifically, we require, as a function of mass: the radius and moment of inertia at TAMS ($R_\mathrm{TAMS}$ and $I_\mathrm{TAMS}$), the maximum angular momenta at He-ZAMS and carbon depletion ($J_\text{He-ZAMS, max}$ and $J_\text{C. dep, max}$), and the squared ratio between the gyration and surface radius $k$. All of these quantities, which are determined by \texttt{MESA} calculations in our work, are metallicity dependent.

\begin{figure}
\includegraphics[width=\columnwidth]{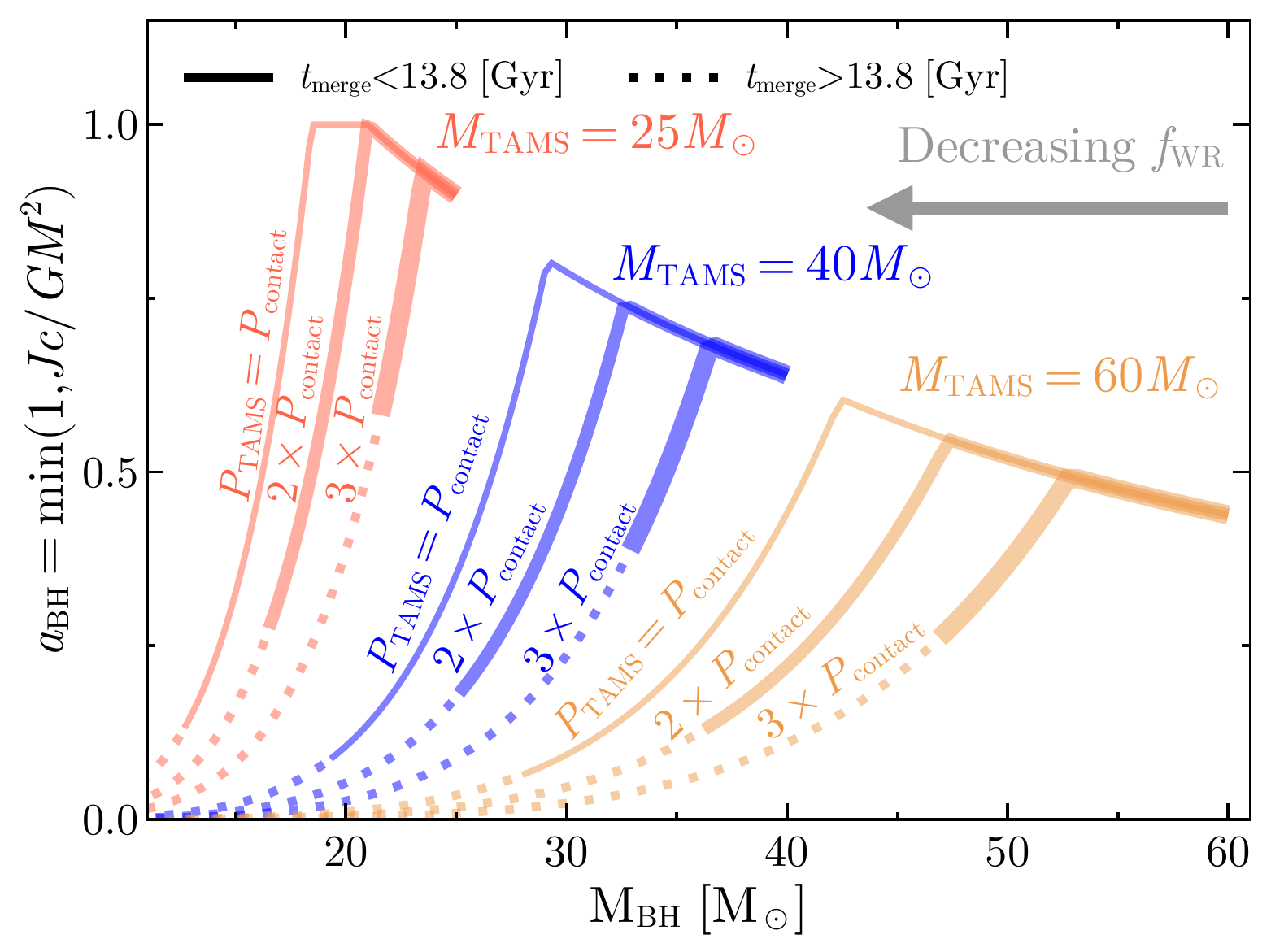}
\caption{Resulting BH masses and spins of CHE stars using a semi-analytical model (see Section \ref{sec:semi}). The model assumes a mass ratio of unity, so two black holes with identical properties are produced. We consider the outcome for three masses at TAMS ($25M_\odot$, $40M_\odot$ and $60M_\odot$) and three orbital periods at TAMS scaled from the period at which the binary would be in contact. Each line represents the results for a fixed $M_\mathrm{TAMS}$ and $P_\mathrm{TAMS}$ with a variable $f_\mathrm{WR}$ (where $M_\mathrm{BH}=f_\mathrm{WR}M_\mathrm{TAMS}$). Cases where the resulting BBH would be too wide to merge within the age of the Universe are marked with a dotted line. Note that for $f_\mathrm{WR}$ close to unity lines with different values of $P_\text{TAMS}$ converge to the same spins at any given mass. The effect of (P)PISNe is ignored in this Figure. Precise boundaries depend on the metallicity used to determine stellar properties in the semi-analytical model used, which in this case where taken from calculations at $Z_\odot/50$.}\label{fig:semievo}
\end{figure}
 
We focus on predictions at the metallicity $Z=Z_\odot/50$ to describe the qualitative results. Results from the other computed metallicities ($Z_\odot/10$ and $Z_\odot/250$) are qualitatively similar. As we find that $k$ covers values between $\sim 0.08-0.1$ during core helium burning, we adopt $k=0.09$. By considering different values for $M_\mathrm{TAMS}$ and $P_\mathrm{TAMS}$ we can explore what sets the highest and lowest spins produced for different BH masses. This is done in Figure \ref{fig:semievo}, where for each choice of $M_\text{TAMS}$ and $P_\text{TAMS}$ we vary continuously $f_\mathrm{WR}$ between unity and zero. The right-most point of each curve represents the case of no mass loss ($f_\mathrm{WR}=1$). The most apparent feature in Figure \ref{fig:semievo} is an exclusion region for high spins and high BH masses. This limit is set by critical rotation at carbon depletion (i.e. the $\Gamma-\Omega$ limit, see Section \ref{test}), and determines the final BH spin as a function of BH mass, independent of the initial orbital period, when WR mass loss is negligible. The $\Gamma-\Omega$ limit is not restricted to CHE evolution, but should be a common feature of evolutionary channels where BHs are formed from helium stars.

More subtle are the conditions that define the minimum spins that can be achieved at any given mass from CHE. There are two ways in which angular momentum could be reduced: either reach TAMS at a higher orbital period (leading to slower rotation due to synchronization), or lose mass to remove angular momentum after TAMS. Since we still need the resulting BBH to merge, we can question which of these two options results in a smaller final spin. Is it better to start at contact at TAMS and then lose as much mass as possible before the orbit widens too much? Or can lower spins be achieved by starting with an orbit at TAMS as wide as possible to yield a merging BBH without any additional mass loss? Figure \ref{fig:semievo} shows that the former is the case. For each $M_\mathrm{TAMS}$ considered, the lowest spin achievable while still resulting in a merging BBH comes from binaries that are at contact in TAMS. An important caveat is that our choice of $f_\mathrm{WR}$ is arbitrary, while in practice the fraction of mass lost will be determined by the metallicity and mass at the He-ZAMS.

\begin{figure}
\includegraphics[width=\columnwidth]{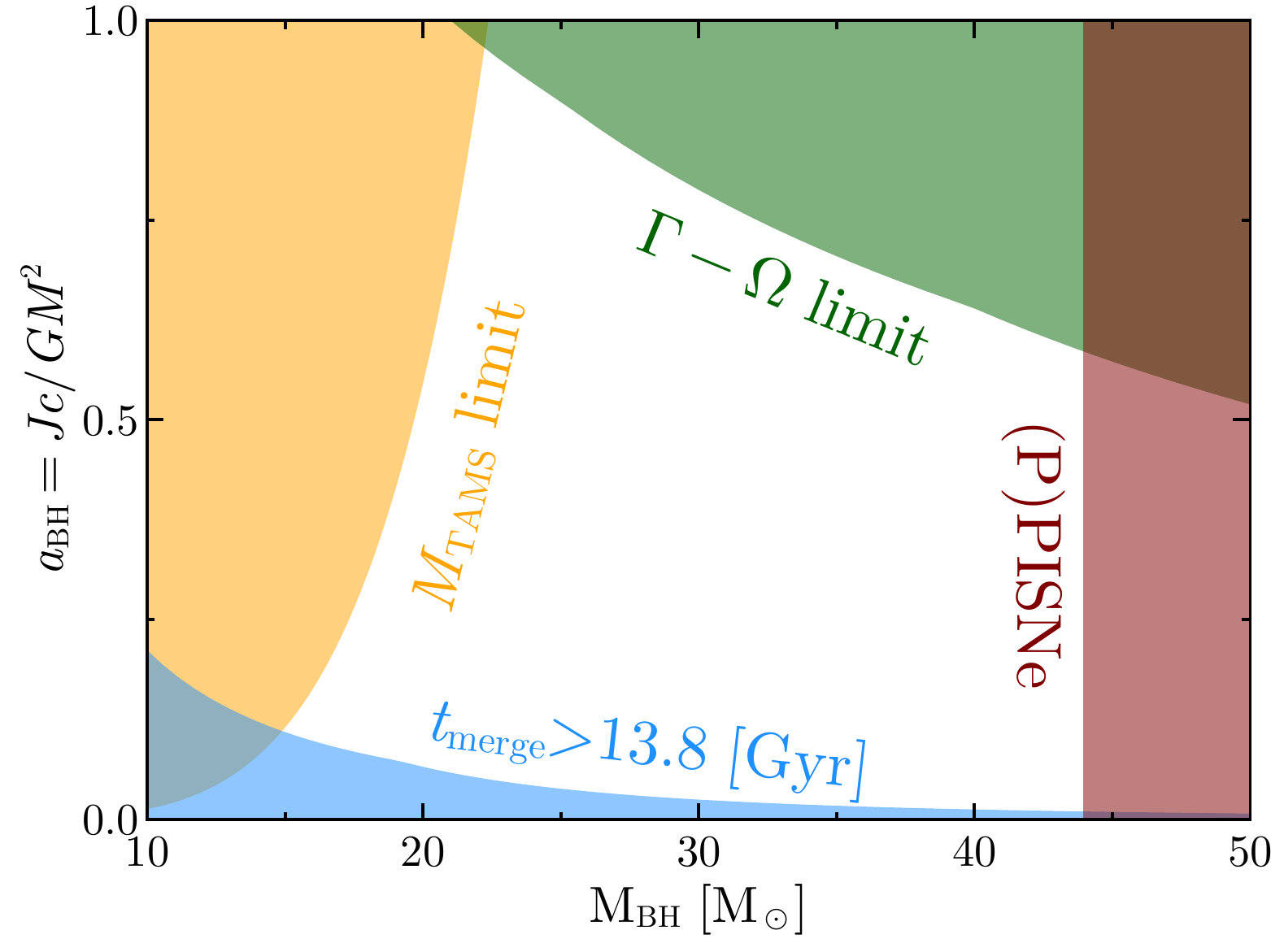}
\caption{Physical processes setting the boundaries for BH masses and spins for the CHE formation scenario. The four colored areas indicate excluded regions due to four different physical processes. In this figure an assumed limit for the $M_\text{TAMS}$ of CHE stars is set at $30M_\odot$, which sets the boundary at the left of the Figure. Precise boundaries depend on the metallicity used to determine stellar properties in the semi-analytical model used, which in this case where taken from calculations at $Z_\odot/50$.}\label{fig:semimodel}
\end{figure}

The upper and lower bounds on the spins discussed so far are illustrated in Figure \ref{fig:semimodel}. Two additional bounds are included, one accounting for the (P)PISNe gap following the results of \cite{Marchant+2019}, and another one coming from the limits of CHE evolution. CHE is expected to be a process favored at higher masses \citep{Maeder1987}, such that we cannot have an arbitrarily low $M_\text{TAMS}$. In figure \ref{fig:semimodel} we illustrate this by drawing an exclusion region that corresponds to $M_\text{TAMS}\leq 30M_\odot$. These boundaries can be compared to the results of detailed simulations to see if this simple model can reproduce the predicted features, which we do in the following section. We point out however that various physical effects are not included in this model. In particular, mass loss is expected to increase at higher masses and metallicities. This also implies that at the high metallicities where we would expect significant post-MS mass loss, mass loss will have potentially widened the orbit of the stars during the MS itself. This restricts the minimum period at TAMS and thus the lower limit on the final spins. Despite this, even in detailed models the impact of mass loss on the orbital period is approximate, and orbital widening can be reduced or even inverted by tidal coupling of the wind \citep{BrookshawTavani1993, macLeodLoeb2020, Schroeder+2021}.

\section{Comparison to population synthesis results}\label{sec:pop}

The semianalytical model presented in the previous Section provides an overall idea of the limits of CHE in terms of masses and effective spins. However, as we vary multiple parameters without accounting for their possible correlation (such as the fraction of mass lost after the MS and the period at TAMS), it is instructive to compare our results with detailed population synthesis calculations. In this Section we re-analyze the population synthesis results of \citetalias{{duBuisson+2020}} by including information on stellar spins and compare them against our results.

\subsection{The population synthesis simulations of \citetalias{duBuisson+2020}}\label{sec:db20}
The work of \citetalias{duBuisson+2020} already provided predictions for the rates and the mass distribution of merging BBHs formed through CHE. These results were based on a large grid of \texttt{MESA} models, exploring the parameter space of initial masses, orbital periods and metallicity. For computational reasons the simulations of \citetalias{duBuisson+2020} were restricted to a mass ratio of $q=1$, which was taken to be representative of the evolution of binaries in the range $q=0.8-1$. In order to provide rate predictions this grid of detailed binary simulations was combined with the theoretical predictions of \cite{TaylorKobayashi2015} for the cosmic star formation history, accounting in particular for the distribution of formed stars at different metallicities. The results of \citetalias{duBuisson+2020} are provided in the form of a comoving box simulation, indicating the formation redshift of all BBHs formed within it through CHE.

\begin{figure}
\includegraphics[width=\columnwidth]{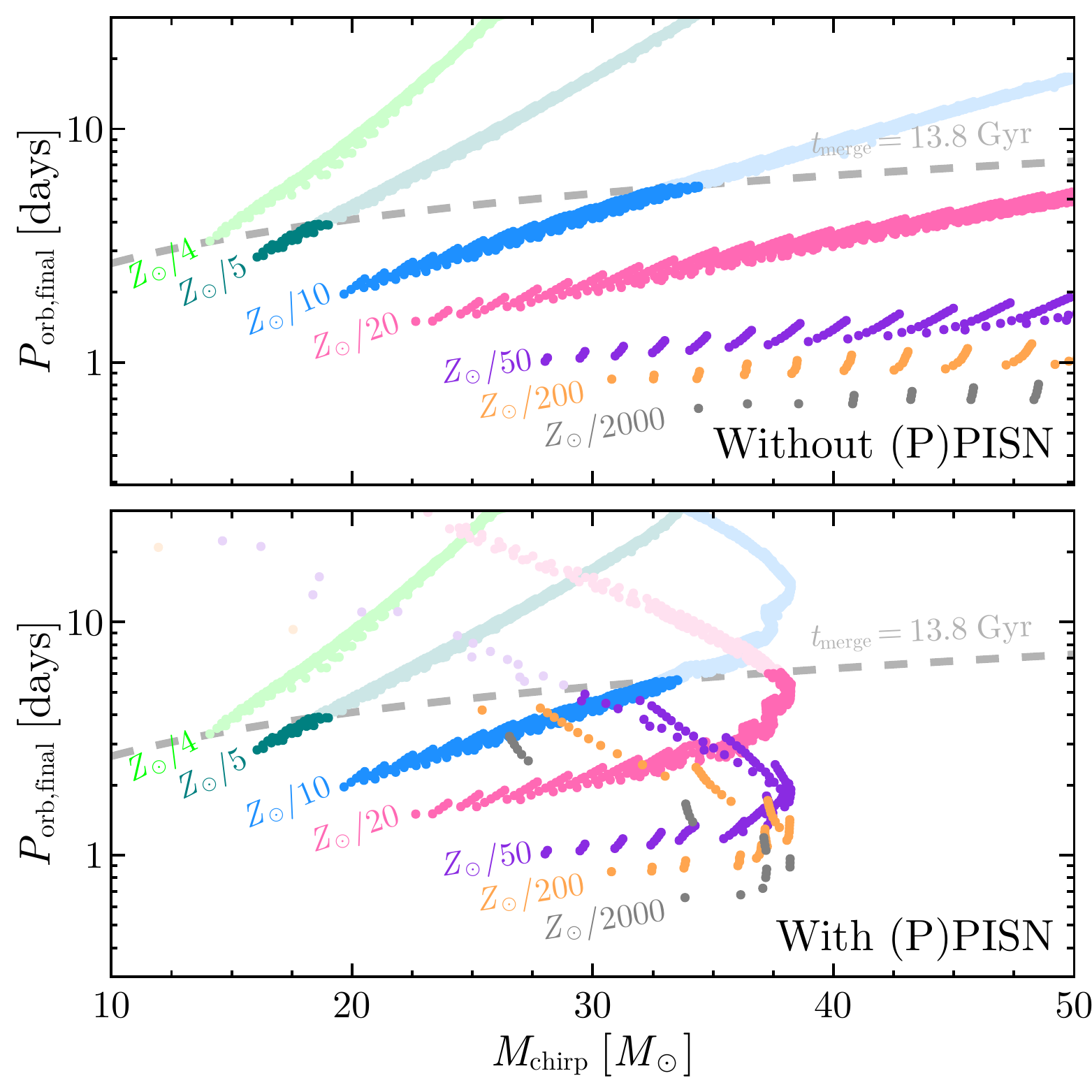}
\caption{(Top) Final orbital period and chirp mass in the \texttt{MESA} simulations of \citetalias{duBuisson+2020}, assuming direct collapse into a BH (with its mass at carbon depletion) and ignoring mass loss due to (P)PISNe. Each dot indicates the final result of one CHE simulation. Systems above the dashed gray line have merger times longer than the age of the Universe and are marked with a lighter color. Results are shown for their grids with metallicities of $\log_{10} Z=-2.375$, $2.5$, $-2.75$, $-3$, $-3.5$, $-4$ and $-5$, with the labels indicating approximately their values in terms of the solar metallicity (taken here to be $Z_\odot=0.017$ as in \cite{Grevesse+1996} for consistency with the definition used in \citetalias{duBuisson+2020}). (Bottom) same as the top panel but after including the impact of (P)PISNe as described in \cite{Marchant+2019}.}\label{fig:MP}
\end{figure}

Consistent with the results of \cite{Marchant+2016}, for any given metallicity the \texttt{MESA} simulations of \citetalias{duBuisson+2020} show a tight correlation between the final orbital period and mass of a binary. This is a consequence of wind mass loss, which at high metallicities lowers the total mass and, under the assumption of Jeans' mode mass loss (stellar winds removing orbital angular momentum equal to that of each component), result in orbital widening. At a fixed metallicity and chirp mass, the small spread in orbital periods is connected to the narrow range in initial orbital periods that are expected to lead to CHE \citep{MandeldeMink2016,Marchant+2016, Song+2016, Riley+2021}. For BH formation through direct collapse, this leads to a similar relationship between the period and the chirp mass of the binary BH (see top panel of Figure \ref{fig:MP}).

Including the effect of (P)PISNe produces a more complex relationship between the chirp mass and the orbital period of BBHs formed through CHE. Mass loss due to PPISNe was included in the results of \citetalias{duBuisson+2020} by interpolating the pre-PPISN to BH mass relationship of \cite{Marchant+2019}. The change in orbital properties was determined by assuming mass loss was instantaneous, followed by circularisation of the orbit at constant orbital angular momentum prior to BH formation. The bottom panel of Figure \ref{fig:MP} shows the resulting final orbital periods and chirp masses after the inclusion of mass loss due to PPISNe. As the relationship between pre-PPISNe mass and BH mass is non-monotonic, the simple relationship between $M_\text{chirp}$ and the final orbital period does not hold anymore. Owing to the ocurrence of PISNe, no BHs are produced between $44M_\odot$ and $123M_\odot$, which constitutes the PISN mass gap according to \cite{Marchant+2019}.

The resulting spins of BBHs formed through CHE were not discussed by \citetalias{duBuisson+2020}. In all simulations that are unaffected by (P)PISNe we have assumed all angular momentum and mass collapse into the BH, resulting in a black hole spin of
\begin{eqnarray}
    a_\text{BH}=\frac{Jc}{M^2G},
\end{eqnarray}
where $J$ and $M$ are the final pre-collapse spin angular momentum and mass. Measurements of individual spins in merging binary BHs have large errors, but the effective spin is much better constrained. As \citetalias{duBuisson+2020} only consider equal mass binaries with aligned spins, $\chi_\mathrm{eff}$ is equal to $a_\text{BH}$ in their simulations. For cases undergoing PPISNe we use the angular momentum profile of the star at the end point of the calculation, which in \citetalias{duBuisson+2020} corresponds to either core-carbon depletion or the onset of collapse due to pair-instability. Taking the final expected BH mass $M_\mathrm{BH}$ from \cite{Marchant+2019} we consider only the angular momentum $J(M_\text{BH})$ contained within the corresponding mass coordinate. The final spin is then computed as
\begin{eqnarray}
    a_\text{BH, PPISN}=\frac{J(M_\text{BH})c}{M_\text{BH}^2G}.
\end{eqnarray}
This assumes the mass ejection from PPISN is instantaneous. One caveat of this approach is that the resulting BH spin can exceed unity, indicating the potential formation of a GRB through the collapsar model \citep{Woosley1993,MacfadyenWoosley1999}. For simplicity we simply cap the dimensionless spin at unity. More detailed models to relate the angular momentum distribution of a stellar model to the final mass and spin of the BH are available (e.g., \citealt{BattaRamirezRuiz+2019}), which consistently produce sub-critical spins.

\begin{figure}
\includegraphics[width=\columnwidth]{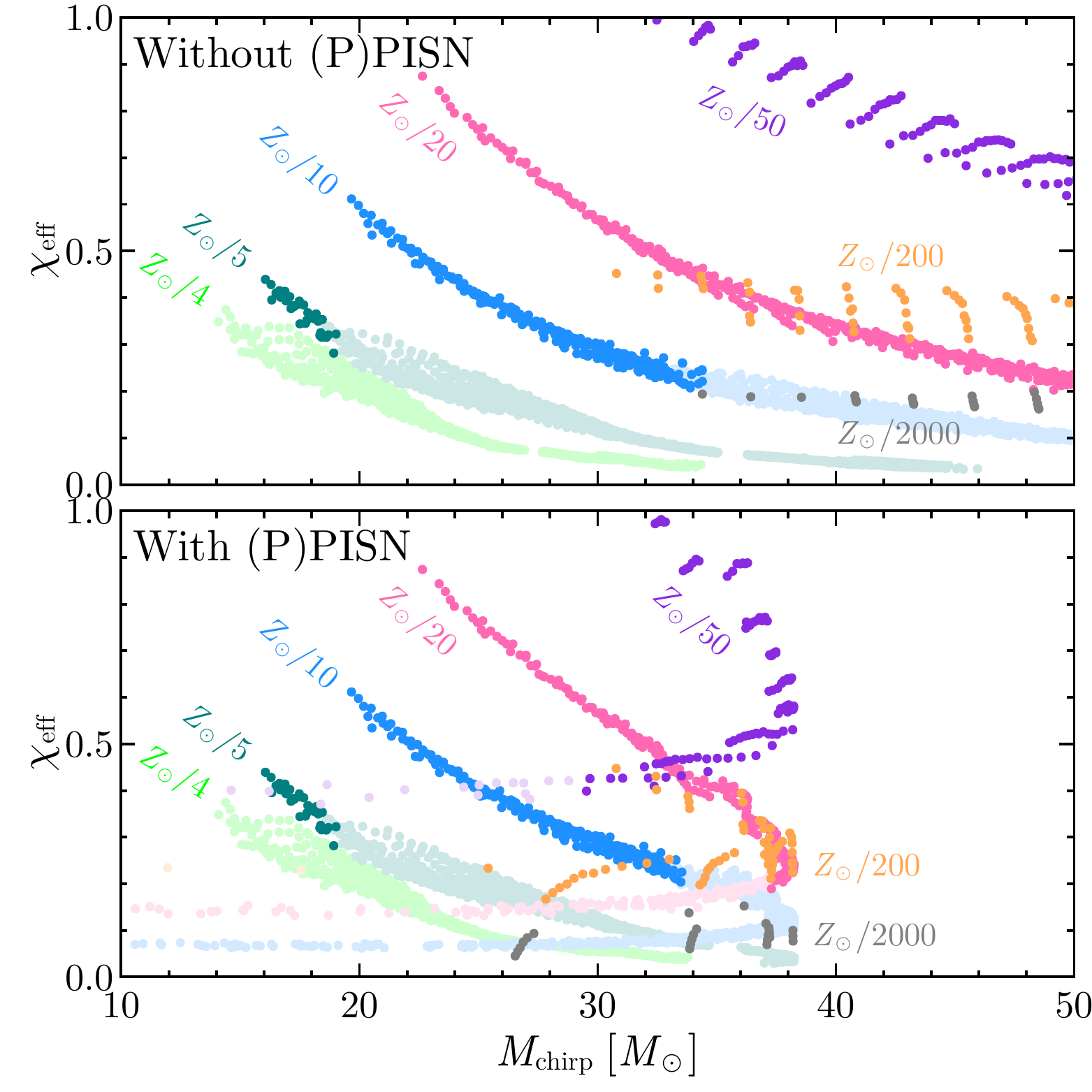}
\caption{Same as Figure \ref{fig:MP} but showing the resulting chirp masses and effective spins of BBHs formed through CHE in the \texttt{MESA} simulations of \citetalias{duBuisson+2020}.}\label{fig:Mchi}
\end{figure}

The resulting chirp masses and effective spins from the \texttt{MESA} models of \citetalias{duBuisson+2020} are shown in Figure \ref{fig:Mchi}. Without the inclusion of (P)PISNe, these two quantities follow clear relationships at each metallicity, with higher chirp masses resulting in lower spins. This is a consequence of wind mass loss after TAMS, lowering the spin of the star before BH formation. At each metallicity the lowest spins produced for merging BBHs correspond to the point where the resulting delay time is equal to the age of the Universe, an effect we explained in Section \ref{sec:semi}. At very low metallicities ($\leq Z_\odot/200$) the \texttt{MESA} simulations produce low spins that break from the general trend. We have identified that in these cases, owing to a smaller relative difference between the size of the stars at TAMS and the He-ZAMS, stars remain synchronous after the end of the main sequence. This leads to angular momentum being removed from each component by tides on the He main sequence, as stars spin up during the contraction phase after core-hydrogen depletion. This effect is potentially spurious, as the model used for tidal evolution is based on the dynamical tide process of \cite{Zahn1977} using pre-computed properties of main-sequence stars (see \citealt{Paxton+2015} for details). Recently \cite{MaFuller2023} have shown that tidal synchronization likely only operates in stripped stars for very short orbital periods, which are forbidden in CHE evolution as the orbit needs to be wide enough to fit the stars during their main sequence evolution.

For this work we have also recomputed the detection probabilities from \citetalias{duBuisson+2020} using updated noise curves. We have used noise curves corresponding to the O3 measured sensitivity, and the O4 and O5/A+ expected sensitivities as described in \citet{Abbott+2020_prospects}. In particular, for O3 we have used the measured noise curve based on three months of data from the LIGO-Hanford detector, while for O4 we have used the pessimistic noise curve estimate with a range for binary neutron star detection of 160 Mpc\footnote{These noise curves are available at \url{https://dcc.ligo.org/LIGO-T2000012/public}}. We have also included the effect of spins on detectability by using the IMPRPhenomB waveform model with non-precessing spins from \cite{Ajith+2011}.

\subsection{Rates of high spin mergers from chemically homogeneous evolution}
\begin{figure}
\includegraphics[width=\columnwidth]{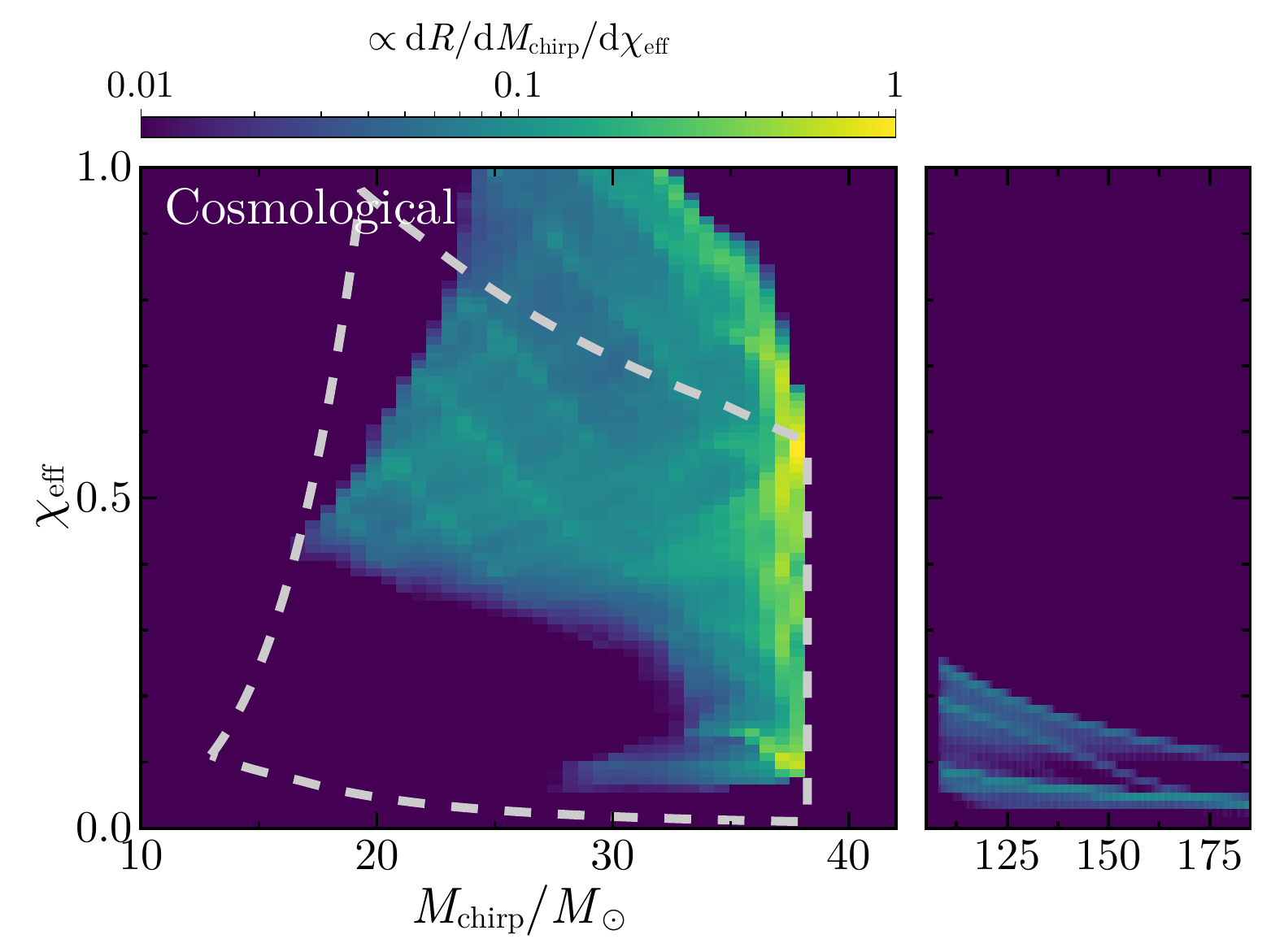}
\caption{Distribution of the rate of merging BBHs per unit chirp mass and unit effective spin that is observable from Earth with a detector with infinite sensitivity, derived from the simulations of \citetalias{duBuisson+2020}. We refer to these rates as ``cosmological''. Numbers are normalized to the maximum value of the differential rate $\text{d}R/\text{d}M_\mathrm{chirp}/\mathrm{d}\chi_\mathrm{eff}$. The two panels highlight predicted systems below and above the PISN gap. The white dashed contour corresponds to the predicted region below the PISN gap that arises from CHE, computed with our semi-analytical model in Figure \ref{fig:semimodel}}\label{fig:cosmo}
\end{figure}

Using the insight developed from the semi-analytical model of Section \ref{sec:semi} we re-analyze the rate predictions of \citetalias{duBuisson+2020} for CHE evolution while considering the predicted spins for these models. Before discussing rates for specific detector sensitivities we first compare the properties of our semi-analytical model with the results of \citetalias{duBuisson+2020} for a detector with infinite sensitivity (which we refer to as ``cosmological'' rates). Figure \ref{fig:cosmo} illustrates how the cosmological BBH mergers are distributed in effective spins and chirp masses. As PPISNe produce a pile-up of systems near the PISN gap \citep{Belczynski+2016b}, we see that the distribution in Figure \ref{fig:cosmo} peaks at $M_\mathrm{chirp}\sim 38M_\odot$. An additional pile-up is seen at the top of the distribution, coming from the spins of BHs being limited by the critical rotation of their progenitor stars. For comparison, we also consider the predicted region for CHE evolution shown in Figure \ref{fig:semimodel}. Although the semi-analytical model qualitatively reproduces various features of the detailed simulations, it is shifted significantly towards lower effective spins. The mergers with $\chi_\text{eff}\lesssim 0.2$ and $30\lesssim M_\mathrm{chirp} \lesssim 40$ correspond to the very metal poor systems that undergo tidal synchronization after TAMS. As discussed in Section \ref{sec:db20} the low spins in these very metal-poor simulations could be spurious.

One of the reasons why the simulations of \citetalias{duBuisson+2020} and our semi-analytical model make a different prediction for the minimum spins produced by CHE was discussed in Section \ref{sec:semi}. The lowest spins predicted arise from binaries that are at contact at TAMS and lose a significant amount of mass during the post-MS. But systems that would undergo significant post-MS mass loss would also be expected to be widened by mass loss before TAMS, so the two free parameters in the semi-analytical model are not really independent. However, another source of discrepancy comes from angular momentum coupling and how close to solid-body rotation the \texttt{MESA} simulations of \citetalias{duBuisson+2020} are. Differential rotation also modifies how the $\Gamma-\Omega$ limit translates into an upper limit on BH spins, as it allows for a larger angular momentum budget in the stellar core without exceeding critical rotation at the surface.

\begin{figure}
\includegraphics[width=\columnwidth]{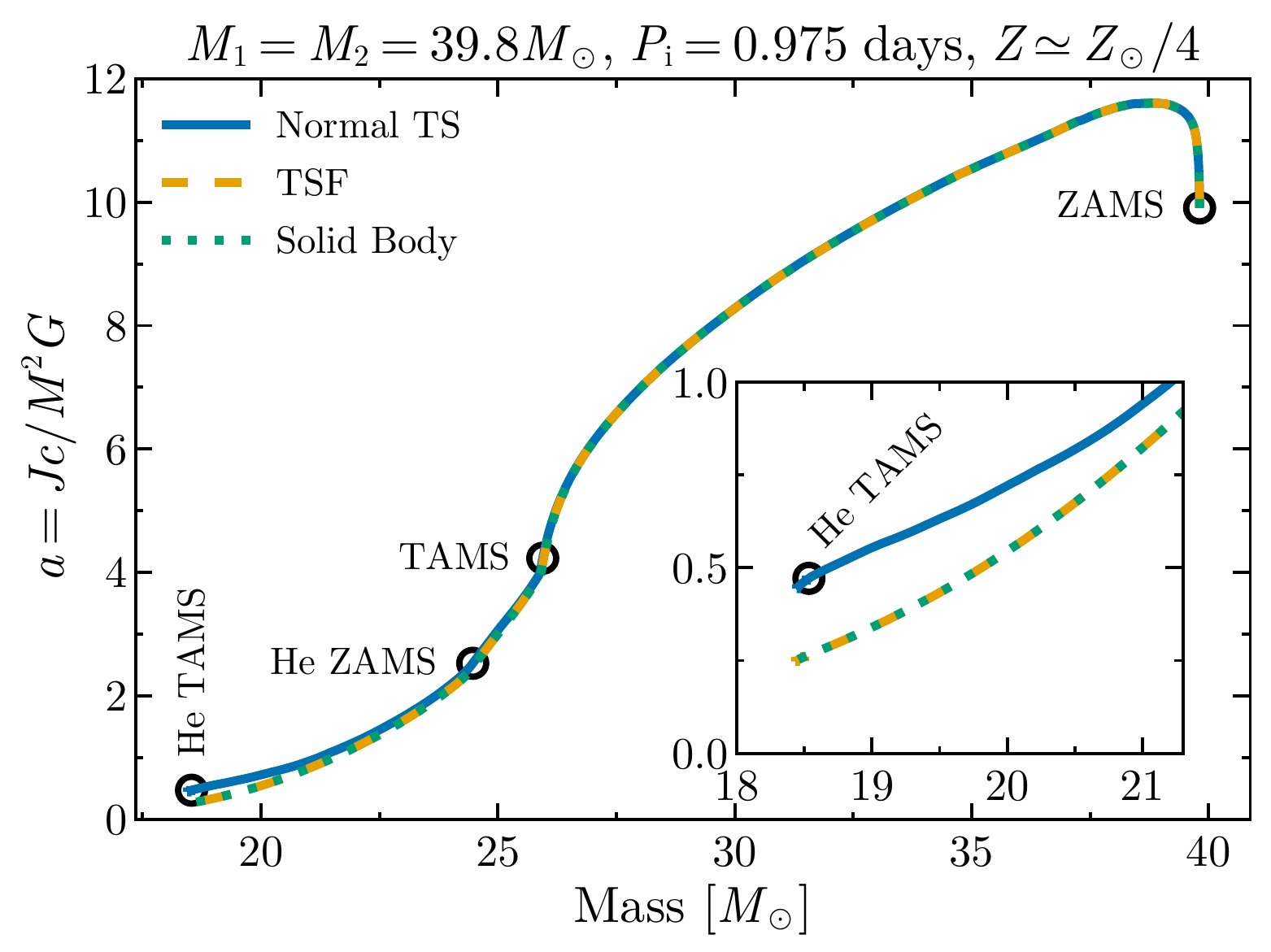}
\caption{Mass of each component versus instantaneous dimensionless spin during the evolution of an equal mass binary undergoing CHE. Shown are the original simulation from \citetalias{duBuisson+2020} using the Tayler-Spruit (normal TS), one recomputed using the modified version of the Tayler-Spruit dynamo from \citet{Fuller+2019} (TSF) and one where solid-body rotation is enforced. These simulations where computed with the same setup and \texttt{MESA} version (11701) used by \citetalias{duBuisson+2020}.}\label{fig:mesasim}
\end{figure}

To test the impact of angular momentum transport in our results, we have recomputed a simulation from \citetalias{duBuisson+2020} where we replaced the Tayler-Spruit dynamo with the modified version from \cite{Fuller+2019}, as implemented by \cite{FullerLu2022}. We also compute a model with an artificially high angular momentum diffusivity in order to enforce solid-body rotation throughout. The simulation chosen corresponds to the merging BBH with the lowest chirp mass produced in the \texttt{MESA} grids of \citetalias{duBuisson+2020}, with $M_\mathrm{chirp}=16M_\odot$ and $\chi_\mathrm{eff}=0.44$. The resulting BH spin from these simulations is shown in Figure \ref{fig:mesasim}. Throughout most of the evolution of the binary, all three simulations match closely, indicating the original calculation was near solid-body rotation. However, the simulations diverege during core-helium burning, with the original calculation from \citetalias{duBuisson+2020} resulting in a much higher black hole spin. The simulation using the modified Tayler-Spruit dynamo of \cite{Fuller+2019} remains almost identical to the one where solid-body rotation is enforced, producing a BBH with $\chi_\mathrm{eff}=0.25$.

Keeping in mind that the results of \citetalias{duBuisson+2020} might overestimate the spins of the resulting merging BBHs, we analyze the range of spins covered by their simulations of CHE. To dissect how different systems contribute, we distinguish the rate between regions of low mass ($M_\mathrm{chirp}<30M_\odot$), high mass ($30M_\odot<M_\mathrm{chirp}<60M_\odot$), and systems above the PISN gap ($M_\mathrm{chirp}>60M_\odot$). Similarly, we distinguish between systems with low ($\chi_\text{eff}<0.5$) and high ($\chi_\text{eff}>0.5$) effective spins. The resulting merger rates are given in Table \ref{table:1}. Focusing on systems below the PISN gap, we find that the dominant contribution ($41\%$) comes from the high-spin and low $M_\mathrm{chirp}$ regime, with high-spin systems constituting the majority of the rate ($61\%$) below the PISN gap. This matches the general view that CHE produces mostly high-spin BHs.

\begin{table*}
\caption{Predicted detection rates (per year of detector time) for BBH mergers produced from CHE, derived from the simulations of \citetalias{duBuisson+2020}. Row marked as "Cosmological" represents the full rate observable from Earth with a detector of infinite sensitivity. Column marked "Full" shows the full rate for each detector, while the other columns represent subsets of the full population filtered by their effective spins and masses.}              
\label{table:1}      
\centering                                      
\begin{tabular}{c  c c c c c c}          
\hline\hline                        
 &  & $\chi_\mathrm{eff}<0.5,$ & $\chi_\mathrm{eff}>0.5,$ & $\chi_\mathrm{eff}<0.5,$ & $\chi_\mathrm{eff}>0.5,$ & \vspace{0.05in}\\    
Detector & Full & $\displaystyle\frac{M_\mathrm{chirp}}{M_\odot}<30$ & $\displaystyle\frac{M_\mathrm{chirp}}{M_\odot}<30$ & $\displaystyle30<\frac{M_\mathrm{chirp}}{M_\odot}<60$ & $\displaystyle30<\frac{M_\mathrm{chirp}}{M_\odot}<60$ & $\displaystyle\frac{M_\mathrm{chirp}}{M_\odot}>60$ \vspace{0.05in}\\    %
\hline                                   
    Cosmological & 8600 & 1200 & 2800 & 1500 & 1300 & 1900 \\      
    O3 & 80 & 31 & 13 & 20 & 7.5 & 7.3 \\
    O4 & 290 & 100 & 61 & 68 & 33 & 30 \\
    O5/A+ & 1700 & 420 & 550 & 320 & 260 & 170 \\
\hline                                             
\end{tabular}
\end{table*}

\begin{figure}
\includegraphics[width=\columnwidth]{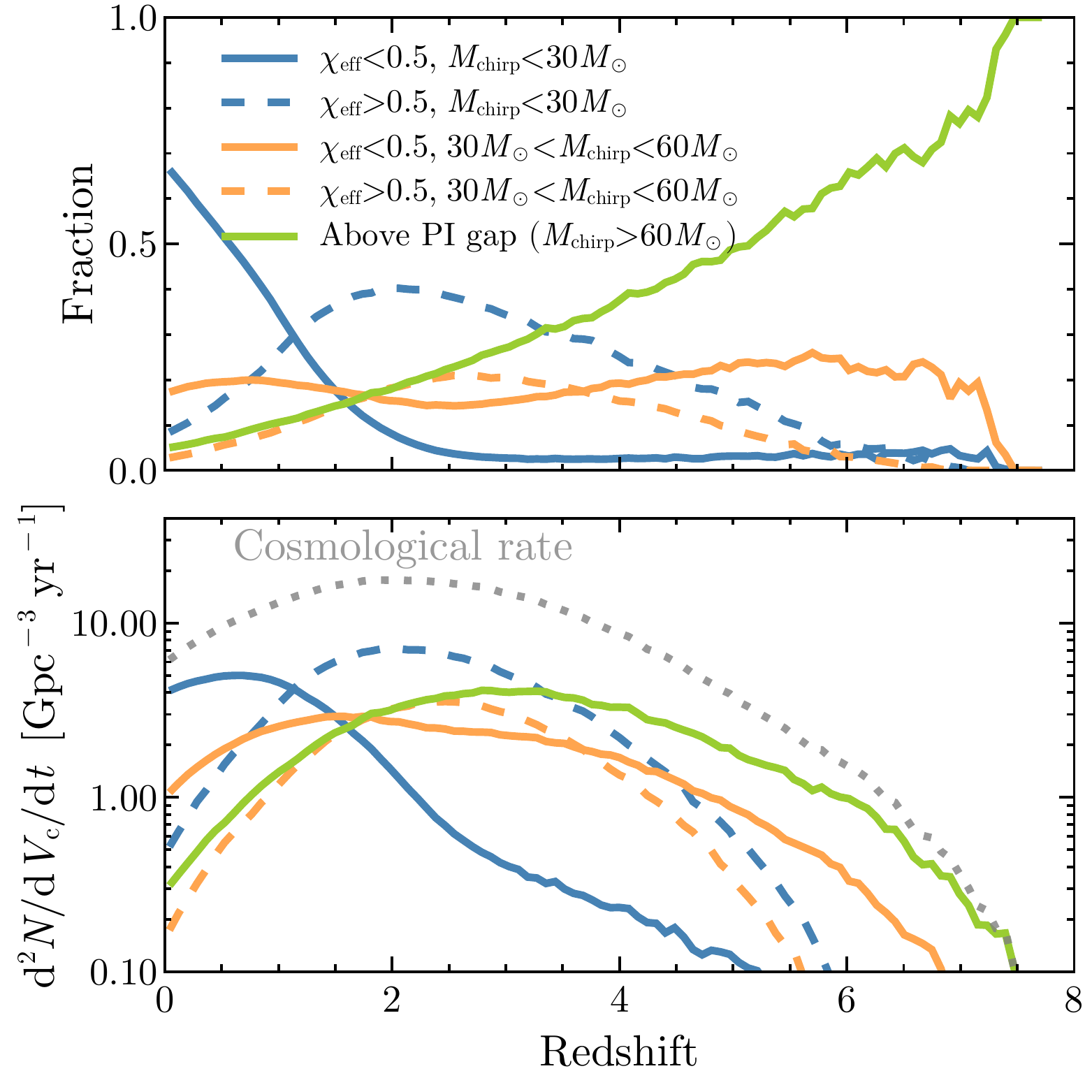}
\caption{Evolution of the BBH merger rate and of different populations of merger events predicted by \citetalias{duBuisson+2020} for CHE. (Top) Fraction of the total rate for specific constraints on the chirp mass and effective spins of BBH mergers. (Bottom) Comoving merger rate corresponding to the same constraints in chirp mass and effective spins. The dotted line indicates the cosmological rate resulting from the addition of all others.}\label{fig:ratez}
\end{figure}

Most high-spin systems come from low metallicity ($Z\lesssim Z_\odot/10$) environments that experience little mass loss and have short merger times (see Figures \ref{fig:MP} and \ref{fig:Mchi}). One way to visualize this is by considering the evolution of the comoving merger rate as a function of redshift, which is done in Figure \ref{fig:ratez}. We can see that at redshift zero the predicted volumetric rate is dominated by low-spin systems below the PISN gap, accounting for 84\% of the local rate. It is only above redshift $1.5$ that high-spin systems constitute the majority of the comoving rate. Observations done with current detectors are not sensitive beyond a redshift of $z \sim 1$, so we expect significant differences between the rate of high-spin mergers in the observable Universe, against those actually observed with current detectors. Another important feature of Figure \ref{fig:ratez} is that systems predicted above the PISN gap become dominant at redshift $\simeq 5$. Predictions above the PISN gap are one of the most uncertain features of the model, as they require the formation of binary stars in excess of $100M_\odot$ for each component \citep{HegerWoosley2002}. However, this makes mergers above the gap an interesting target for 3rd generation detectors which will be sensitive to such massive mergers at redshifts beyond 10 \citep{HallEvans2019}.

\begin{figure}
\includegraphics[width=\columnwidth]{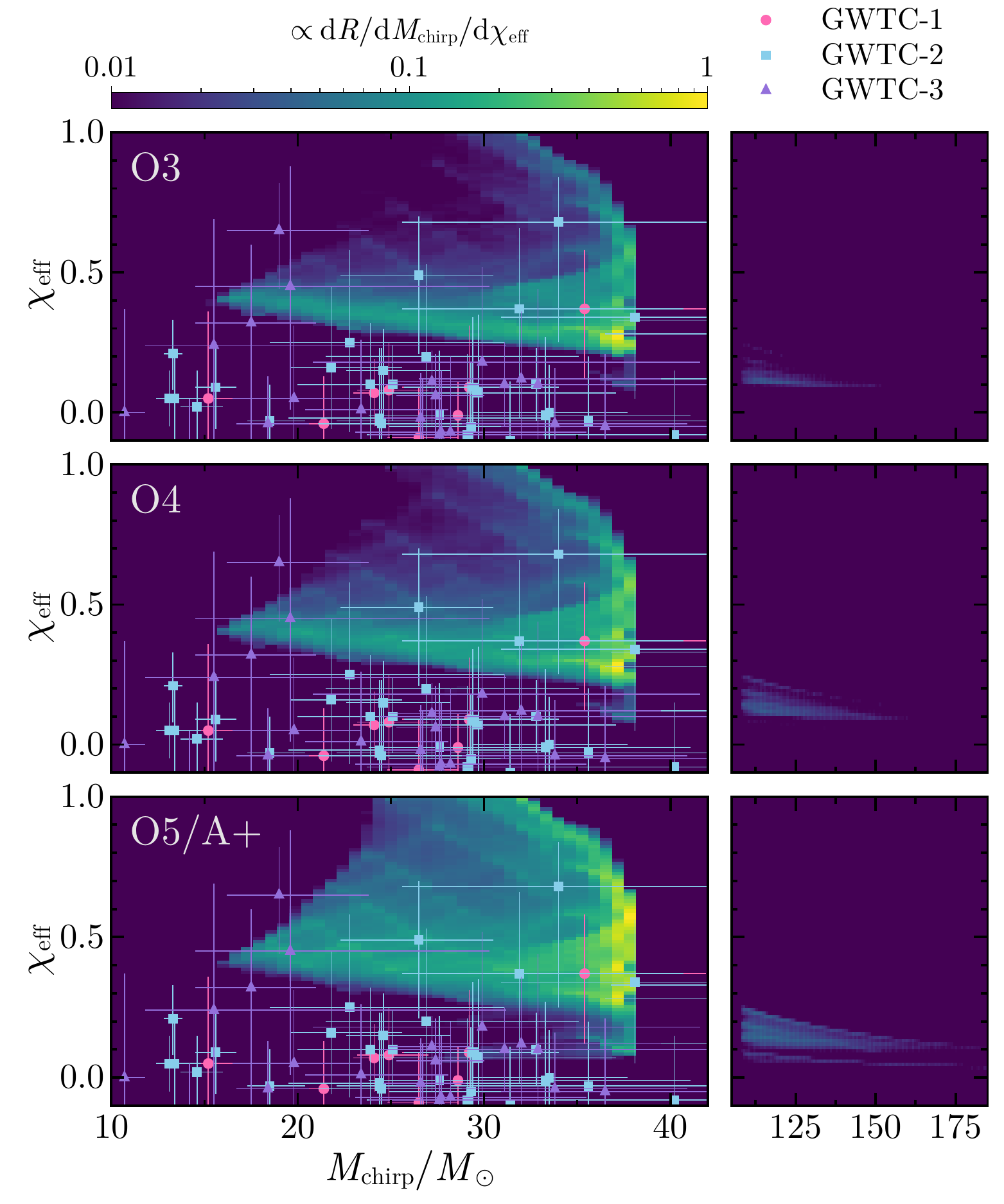}
\caption{Observable rate of BBH mergers considering the detector sensitivity of O3, O4 and O5 for the LIGO detectors. The distribution of the rate is shown as a function of $M_\mathrm{chirp}$ and $\chi_\text{eff}$. Data points indicate all observations made so far as reported in the current gravitational wave transient catalogues (GWTC-1, GWTC-2 and GWTC-3). For each observation the median values of $M_\mathrm{chirp}$ and $\chi_\mathrm{eff}$ are shown together with their 90\% credible intervals.}\label{fig:ratedist}
\end{figure}

As current ground based GW detectors are not sensitive to high redshift mergers, we expect the observed population of BBHs formed through CHE to be dominated by low-spin systems. Figure \ref{fig:ratedist} shows the detectable rate distribution as a function of $M_\mathrm{chirp}$ and $\chi_\text{eff}$ with the measured sensitivity for O3 as well as projected sensitivities for O4 and O5 (which is planned to include the A+ upgrade). Owing to the lack of sensitivity at redshifts above unity, both for O3 and O4 we find that the majority of observed BBHs would populate a narrow band going from $M_\mathrm{chirp}\simeq 18M_\odot$ and $\chi_\mathrm{eff}\simeq 0.45$ down to $M_\mathrm{chirp}\simeq38$ and $\chi_\mathrm{eff}\simeq 0.25$. This corresponds to the limit for which, at any given metallicity, BBHs that are too wide to merge in the age of the Universe are produced. Compared to the currently observed population we find that this band lies at higher $\chi_\mathrm{eff}$ than the majority of observed sources. As shown in Table \ref{table:1}, for the sensitivity of O3 we would expect 30\% of observed sources produced by CHE below the PISN gap to have $\chi_\mathrm{eff}>0.5$, while only a single source has been detected with both median component masses below $60M_\odot$ and a median effective spin above $0.5$ (GW200308\_173609). The predicted $\chi_\mathrm{eff}$ band of observable CHE mergers would be lowered by about a half if we assumed the enhanced Tayler-Spruit dynamo of \citet{Fuller+2019}.

As the sensitivity of detectors increases in the future, the contribution from highly-spinning mergers produced through CHE should become more important in observations. As shown in Figure \ref{fig:ratedist} and Table \ref{table:1}, high-spin ($\chi_\text{eff}>0.5$) and low-spin ($\chi_\text{eff}>0.5$) CHE BBHs are predicted make similar contributions to observations during the fifth observing run, with 52\% of mergers detected below the PISN gap corresponding to the high-spin population. This indicates that direct constraints on the evolution of the merger rate as a function of redshift (including subpopulations of merging BBHs) will be possible before the onset of third-generation detections. For instance, current observations show an indication that the distribution of $\chi_\mathrm{eff}$ broadens with redshift \citep{Biscoveanu+2022}.

Our predicted observable rates for merging BHs formed by CHE would require that almost all GW sources observed to date are a consequence of CHE (see also \citealt{StevensonClarke2022}).
Predicted rates are also in direct contradiction for mergers with both components above the PISN gap, which have not been observed so far. The third observing run of the LVK collaboration, for which detections are described in GWTC-2 and GWTC-3 \citep{GWTC2,GWTC3}, accumulated a total of 275 days of observations with at least two detectors. According to our results for O3, this would translate into $5.5$ detections of mergers above the gap. The lack of such detections could point to weaknesses in the CHE model, but also in our understanding of mass loss rates at very low metallicities and of the distribution of orbital properties of very massive binaries. However, we do find that BHs formed above the gap should have spins $a<0.25$, owing to the limit set on their progenitors by critical rotation. This should be the case for any BH formed above the gap from a stripped-envelope star (not just through CHE) and should serve to clearly distinguish them from objects formed through hierarchical mergers, which are expected to have spins $a\gtrsim 0.7$ \citep{Pretorius2005,BertiVolonteri2008,GerosaBerti2017}.

\section{Conclusions}\label{sec:conclusion}
In this work we describe an upper limit on the spins of merging BBHs formed through binary evolution. This limit arises from the critical rotation rate of stripped star progenitors before their collapse, preventing the formation of BHs with spins near unity for BH masses above $\gtrsim 25M_\odot$.
To study how this limit arises through binary evolution we have considered BBH formation through CHE, a process that allows for the formation of rapidly rotating BH progenitors. We have analyzed the BH spins produced by CHE with a semianalytical model, as well as by reanalyzing the population synthesis results of \citetalias{duBuisson+2020} by including information on BH spins formed through CHE. We compare our results to current observations and provide predictions for upcoming observing runs of ground-based GW detectors.

Based on our semi-analytical model, we identified three different processes (in addition to the PISNe gap) that restrict the range of masses and effective spins that can be achieved through CHE (see Figure \ref{fig:semimodel}). The lowest achievable spins come from CHE binaries with delay times close to the age of the Universe, making this population favored in current observations as they merge at low redshifts. Another restriction is set by the minimum mass for CHE to operate, resulting in a boundary with increasing $\chi_\text{eff}$ as a function of $M_\text{chirp}$. Finally, we find that CHE binaries that experience little mass loss after the main sequence have spins determined by the critical rotation rate of their progenitors, which is determined by a combination of gravity and radiation pressure for massive, luminous stars. Identifying any of these features in GW observations would provide important insight into stellar evolution processes that remain poorly constrained. Although we have studied only the evolution of BBH progenitors through CHE, the spin limit due to critically rotating BH progenitors should be shared by any formation channel for BBHs that involves the collapse of stars without a hydrogen envelope.

Comparing our analytical models with the simulations of \citetalias{duBuisson+2020} we find that, although qualitatively the expected boundaries in $M_\text{chirp}$ and $\chi_\text{eff}$ are present, the final spins are higher in the detailed binary simulations. In part we attribute this to angular momentum transport, as simulations with more efficient coupling between their cores and outer layers can lower the final spins by a factor of $\sim 2$. Coupling is expected to impact both the lower and upper limits of spins that can be produced by CHE, with stronger coupling resulting in lower final spins owing to increased loss of angular momentum through stellar winds. The simulations of \citetalias{duBuisson+2020} predict that during the ongoing O4 run (and also previous observing runs) the majority of detected BBHs formed through CHE would have effective spins smaller than $0.5$.  If the angular momentum coupling is strong, the majority of sources should have spins below $\sim0.25$. Only in future science runs will the high-spin population (which merges promptly at high redshifts) become dominant.

The upper limit on BH spins we have identified also has consequences for the formation of gamma-ray bursts from collapsing stars through the collapsar model \citep{Woosley1993,MacfadyenWoosley1999}. If BHs with spins near unity are required to produce a highly relativistic jet through the Blanford-Znajek process \citep{BlanfordZnajek1977,McKinney2005}, then the progenitors of lGRBs formed by stripped stars would be limited to pre-collapse masses $\lesssim 20M_\odot$. Only broad-lined Ic supernovae have been associated with lGRB events from massive star collapses \citep{Galama1998}, possibly indicating that only stripped stars are their progenitors. The precise limits on progenitor masses are sensitive to angular momentum coupling, and future GW observations can help to identify it. Constraining which stellar progenitors can produce lGRBs is also critical to link observations of high-spin merging BBHs with the observed rates of lGRBs \citep{Bavera+2022}.

Finally, a detection with both BHs unambiguously above the PISN gap is still missing and in contradiction with the results of \citetalias{duBuisson+2020}. This could be related to our lack of knowledge of the distribution of binary properties at both very high masses and low metallicities. Perhaps the stellar binaries we predict to form merging BBHs above the gap do not exist in nature. Another big uncertainty concerns the location of the PISN mass gap which depends strongly on the value of the $^{12}\text{C}(\alpha,\gamma)^{16}\text{O}$ reaction rate \citep{Takahashi2018, Farmer+2019}. Recent calculations using updated values for this rate \citep{Mehta+2022,Farag+2022} have pushed the mass gap to the range $\sim 60M_\odot -140M_\odot$ in BH masses, shifting both boundaries upwards by $\sim 20M_\odot$. This can further lower predicted rates for BBHs above the PISN gap with a corresponding increase in the amount of detections below it. In any case, we expect the spins of BHs formed above the gap to be significantly restricted by the limit of critical rotation on their progenitors, with spins below $0.25$, providing a clear distinction from BHs formed through hierarchical mergers.

\begin{acknowledgements}
PM acknowledges support from the FWO junior fellowship number 12ZY520N and the FWO senior fellowship number 12ZY523N. IM is a recipient of the Australian Research Council Future Fellowship FT190100574.  Analysis of results relied on the \textit{Julia} programming language \citep{Julia-2017} and the \textit{Makie} plotting package \citep{Makie}.
\end{acknowledgements}

\bibliographystyle{aa}

\appendix

\section{Details on microphysical inputs of \texttt{MESA} simulations}\label{appendix::mesa}
Our \texttt{MESA} calculations rely on various inputs regarding microphysical processes, which we list here. Opacities for various temperature and density regimes are taken from OPAL \citep{IglesiasRogers1996}, low-temperature opacities from \cite{Ferguson+2005} and Compton-scattering opacities from \cite{Poutanen2017}. The equation of state is determined from a combination of \citet{RogersNayfonov2002}, \cite{Irwin2012}, \citet{Saumon+1995}, \citet{Timmes+2000}, \citet{PotekhinChabrier2010} and \citet{Jermyn+2021}, with different equations of state used at different conditions for temperature, density and composition (see \citealt{Jermyn+2023} for details on how these are patched together). Nuclear reaction rates are taken from \citet{Angulo+1999} and \citet{Cyburt+2010}, with screening accounted for as in \citet{Chugunov+2007}. Thermal neutrino emission is determined as in \citet{Itoh+1996}.

\section{Fit to maximum BH spin} \label{app:spin}
The values of the maximum dimensionless spin derived from our simulation are given in Table \ref{table:2}. These correspond to the value of $a_\mathrm{max}$ at carbon depletion in Figure \ref{fig:maxspin}. For convenience, we provide the following fit to our results:
\begin{eqnarray}
    a_\mathrm{max}(x) = 10^{c_0 + c_1x+c_2 x^2},
\end{eqnarray}
where $x=\log_{10}(M_\mathrm{BH}/M_\odot)$. The dependence on metallicity is included through $c_0$, $c_1$ and $c_2$,
\begin{eqnarray}
c_0 &=&0.13\sqrt{z} \\
c_1 &=&0.74-0.19\sqrt{z} \\
c_2 &=&-0.58 + 0.12 \sqrt{z},
\end{eqnarray}
with $z=\log_{10}(0.1Z_\odot/Z)$. This fit has not been tested beyond the range of masses and metallicities explored in Section \ref{test}. It is also important to point out that these predictions rely on an understanding of the radii of very massive helium stars near the Eddington limit. Our results have been computed using hydrostatic stellar evolution models, but an accurate prediction of the stellar radius in this regime requires the inclusion of hydrodynamics (e.g. \citealt{Poniatowski+2021}).

\begin{table}
\caption{Maximum black hole dimensionless spin $a_\mathrm{max}$ as a function of black hole mass and progenitor metallicity. These results correspond to the values at carbon depletion of Figure \ref{fig:maxspin}.}              
\label{table:2}      
\centering                                      
\begin{tabular}{c c c c}          
\hline\hline                        
& \multicolumn{3}{c}{$a_\mathrm{max}$}\\
$\log_{10}\left(M_\mathrm{BH}/M_\odot\right)$ & $Z_\odot/10$ & $Z_\odot/50$ & $Z_\odot/250$ \\ 
\hline                                   
0.8 &1.74 & 1.97 & 2.02 \\
1.0 & 1.35 & 1.55 & 1.60 \\
1.2 & 1.03 & 1.20 & 1.25 \\
1.4 & 0.753 & 0.905 & 0.945 \\
1.6 & 0.517 & 0.649 & 0.684 \\
1.8 & 0.301 & 0.423 & 0.453 \\
2.0 & 0.139 & 0.240 & 0.269 \\
2.2 & 0.0583 & 0.121 & 0.148 \\
2.4 & 0.0271 & 0.0514 & 0.0619\\
\hline                                             
\end{tabular}
\end{table}
\end{document}